%% file: conference_101719.tex
\documentclass[conference]{IEEEtran}
\IEEEoverridecommandlockouts
\usepackage{amsmath,amssymb,amsfonts}
\usepackage{algorithmic}
\usepackage{graphicx}
\usepackage{textcomp}
\usepackage{xcolor}
\usepackage[acronym]{glossaries}
\usepackage[numbers]{natbib}       
\usepackage{float}
\usepackage{caption}
\usepackage{subcaption}
\usepackage{hyperref}
\usepackage{booktabs}

\loadglsentries{acronyms}
\def\BibTeX{{\rm B\kern-.05em{\sc i\kern-.025em b}\kern-.08em
    T\kern-.1667em\lower.7ex\hbox{E}\kern-.125emX}}
\begin{document}

\title{Haptic Interactions for Extended Reality\\
\thanks{This work was funded in part by the Research Foundation – Flanders (FWO), IDLab (Ghent University – imec), Flanders Innovation and Entrepreneurship (VLAIO), and the European Union.}
}


\author{
	\IEEEauthorblockN{
		Yentl Vermeulen$^1$, Sam Van Damme$^1$, Glenn Van Wallendael$^1$, Filip De Turck$^1$, \\
		Maria Torres Vega$^{1, 2}$
	}
	\IEEEauthorblockA{
		$^1$IDLab, Department of Information Technology (INTEC), Ghent University - imec, Belgium \\
		$^2$eMedia Research Lab, KU Leuven, Belgium
	}
}

\maketitle

\begin{abstract}
This research investigates whether the interaction methods of \acrfull{xr} headsets can be improved by using haptic feedback. As a first and most common technique, indirect interactions are considered. Indirect interactions correspond to manipulations of virtual objects from a virtual distance using pre-defined hand gestures. As a second interaction technique, direct interaction (namely \acrfull{dim}) has been implemented where the user manipulates objects by virtually touching these with their hands. A third interaction method extends the previous one with haptic feedback (namely \acrfull{hedim}). These 3 methods are compared with each other based on objective and subjective user tests, also taking into account financial considerations. This research concludes that the \acrshort{dim} improves upon the standard indirect method. Additionally, it has been observed that haptic feedback could enhance the \acrshort{dim} in specific situations. Nevertheless, when considering the current financial cost, our subjects were not convinced of the small improvements haptic feedback brings. 
\end{abstract}


\section{Introduction}
Current immersive systems lack tactile feedback when interacting with computer generated objects. Users do not know when they are interacting with a virtual object in \acrfull{xr} because they do not experience any feedback. Additionally, immersive systems have no standard mechanism for interaction comparable to what the computer mouse represents for the personal computer. The computer mouse as an interface led to a breakthrough with the general public and this is what \acrshort{xr} is still waiting for \cite{source_110}. Furthermore, the absence of a direct interaction method is unnatural from the user’s perspective \cite{source_111}. 

Since the interaction with an \acrshort{xr} environment varies from one device to the other, this study considers the HoloLens (first edition) and the Oculus Quest 2 as \acrfull{hmd}. The interaction method of the Hololens will serve as a baseline. 

Currently, there are three methods to operate the HoloLens. Firstly, there is gazing for selecting an item. A gaze is made when the user moves the whole device, including their head, and try to aim the projected pointer on the desired item. Secondly, the air tap is used to interact with the object on which the pointer is projected. To perform an air tap, the user taps their thumb and index finger together in sight of the camera. The third and final option is a voice command, which is not standard integrated for all applications. 

This research will focus on the integration of two new interaction methods. Firstly, the \acrfull{dim} is proposed to make the standard interaction method more natural. Secondly, this new interaction method is extended with the Prime II Haptic Edition gloves from Manus and is consequently named the \acrfull{hedim}. 

\begin{figure}[htbp]
\centerline{\includegraphics[width=0.5\textwidth]{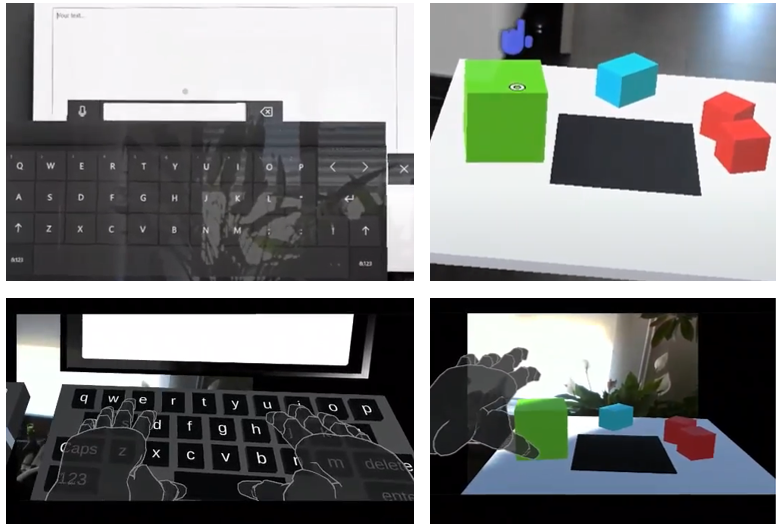}}
\caption{Visualization of the tasks to be performed by the subjects. Assignments on the HoloLens (top row) and assignments on the Oculus (bottom row).}
\label{fig:diagram}
\end{figure}

The purpose of this research is to examine if these new interaction methods improve the user experience, accuracy, and speed of performing frequently used \acrshort{xr} tasks in everyday life. As assignments, it is chosen to let the users type a sentence on a virtual keyboard and construct a tower of virtual cubes. Both assignments used in this research are visualized in Figure \ref{fig:diagram}. After each set of tests, the user will receive a questionnaire with subjective questions about their preference of use, user experience, cybersickness and possible irritations. 

\section{Taxonomy}

\begin{figure}[htbp]
\centerline{\includegraphics[width=0.5\textwidth]{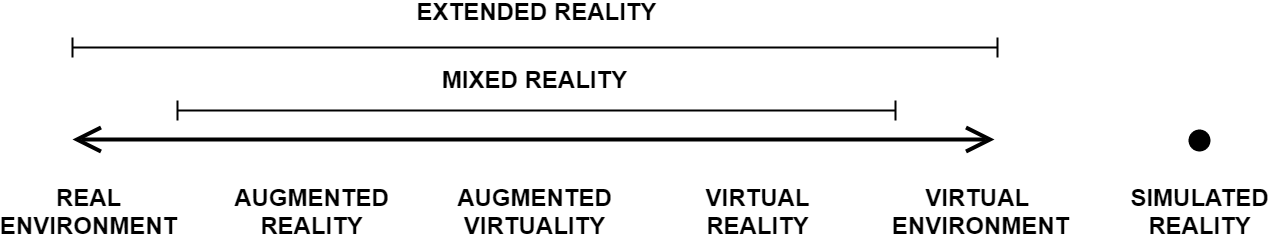}}
\caption{Updated version of the reality-virtuality continuum.}
\label{fig:reality-virtuality_continuum}
\end{figure}

To make it easier to understand how XR works, it is essential to know what it refers to. Therefore, all terms used will be visualized and defined based on the reality-virtuality continuum. This continuum is a continuous scale ranging from the real environment to the completely virtual environment. It encompasses all possible variations and compositions of real and virtual objects and was first introduced in the research of Milgram and Kishino in 1994 \cite{source_51}. The first version did not contain all contemporary terms; therefore, an update is performed as shown in Figure \ref{fig:reality-virtuality_continuum}. This study will only consider the descriptions below as definitions to avoid misunderstandings, as some terms are not strictly defined and are used interchangeably in different contexts.

\begin{itemize}

    \item EXtended reality (\acrshort{xr}) is the term referring to all real-and-virtual combined environments and human-machine interactions generated by computer technology and wearables, where the X represents a variable for any current or future spatial computing technologies \cite{source_98}. \acrshort{xr} can be understood as an umbrella term that includes the entire reality–virtuality continuum and its extremes.  

    \item \acrfull{mr} or hybrid reality is a subset of \acrshort{xr} and refers to everything in the continuum except for all applications on the two extremes \cite{source_51}. All applications between the two extremes can be described as a combination of the real and the virtual world, meaning that both realities are mixed. In non-academic context, \acrshort{mr} is often referred to as an extra step of \acrshort{ar} where holograms not just overlap the real world but in addition, they are interactive and manipulable \cite{source_99}. 
    
    \item \acrfull{ar} is part of \acrshort{mr} and lies between the two extremes. It refers to an enhanced version of reality created by using technology to overlay digital information on an image of something being viewed through a device \cite{source_51}. To retain the essential components and avoid limiting \acrshort{ar} to specific technologies three characteristics are often used, namely, it is a combination of real and virtual content, it is interactive in real time and it is registered in 3D \cite{source_20}.

\end{itemize}

\section{State of the art}
Currently, \acrshort{hmd}-companies use their own created interaction method for their product. This creates confusion among users when a new product is purchased. Additionally, their interaction method is never investigated in the public scientific domain or compared with other methods to define the best possible solution.

Despite the possibilities for daily use such as \acrshort{ar} meetings with Microsoft Mesh, users prefer to a personal computer for their daily use \cite{source_94}. This research will not compare the current interaction method of XR-devices with a personal computer, but it will try to enhance the basic interactions used by XR-devices based on common daily applications with an ultimate goal of bringing XR-interactions closer to traditional PC interactions from a productivity point of view. 

Haptic feedback has been investigated in a lot of contexts~\cite{overview}. 
Starting at the hardware of the interface, it can be provided originating from different sources, such as straight from the display device upon touching it~\cite{HoloHaptics}, but the focus of this work is rather on gloves worn during an AR-interaction. 
When considering wearable haptic interfaces, there can be a focus on different purposes of the haptic feedback, such as identifying objects~\cite{3DObjectIdentifyHaptics} or productivity as envisioned in this work. 
Looking from a different angle, haptic interfaces can be used to simulate working with real-life tools~\cite{UXHapticsVR}~\cite{HapticToolDrill}. Here, the difference is more subtle, but although we consider a keyboard and cubes in our test which correspond to real life objects, it is not our goal to make these feel like the real counterparts. 
In contrast to previous work~\cite{3dof6dofcomplexhaptic} where the interaction capabilities of different people are compared to each other in a game like fashion, this study tries to quantify individual differences in productivity when utilizing different interfaces.
Finally, no studies regarding the interest in buying an \acrshort{hmd} and haptic feedback were found. Consequently, this research investigates how much participants want to spend on an \acrshort{hmd} and whether they can estimate the value of the used \acrshort{hmd}s.

\section{Proposed solution}
Because this research focuses on the basic interactions, the assignments for the subjects are deliberately kept simple. Devising complex tests where the user is pushed to the limit and is given a big explanation would only negatively influence the results. That is why it was decided to perform a simple keyboard task and building a virtual tower of blocks. These are tasks that are straight forward for any user at first glance and require little or no practice. Additionally, these assignments have the advantage that both the three-dimensional and the two-dimensional space are tested. 

\subsection{System architecture}
\begin{figure}[htbp]
\centerline{\includegraphics[width=0.5\textwidth]{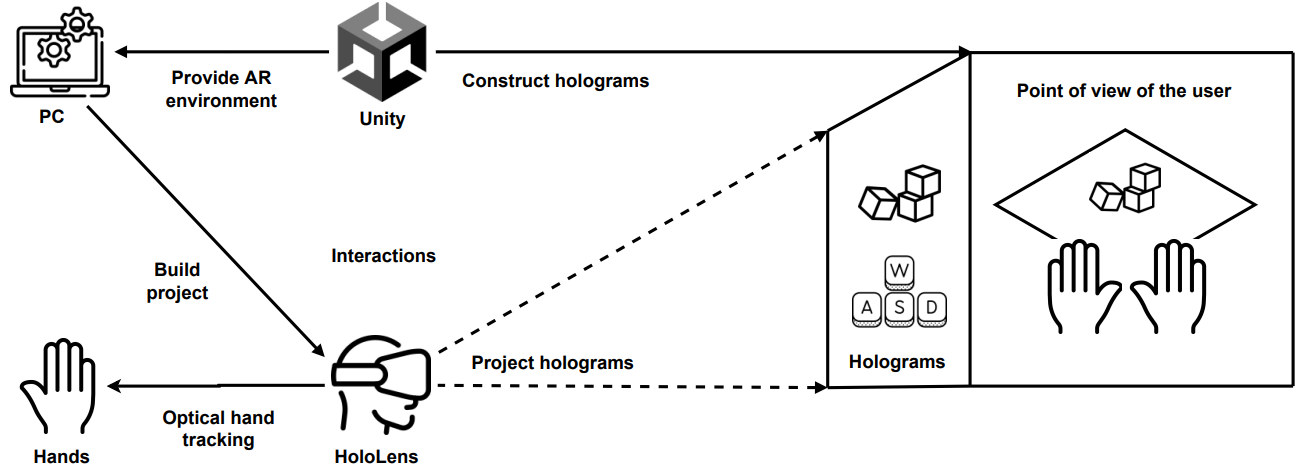}}
\caption{Schematic visualization of the system architecture using the standard interaction method of the HoloLens.}
\label{fig:interactie1}
\end{figure}

Firstly, a schematic visualization of the system architecture using the standard interaction method of the HoloLens is shown in Fig.~\ref{fig:interactie1}. On the right in this image, the point of view of the user can be seen. It shows a couple of cubes that can be constructed into a tower. Similarly, a virtual keyboard can be placed in the \acrshort{ar} environment. This environment is constructed using the Unity game engine, which is installed on a personal computer. When an \acrshort{ar} environment is coded, it can be tested on the HoloLens. This can be done by streaming the project or let it build completely standalone on the \acrshort{hmd}. Due to it can only be streamed wirelessly, streaming is only used for testing purposes. If a user is testing the standalone version of an assignment on the HoloLens, they can interact with the holograms using the standard HoloLens interaction methods, such as gaze and air tap. In Unity, the HoloLens is coded to perceive the hands of the user via the built-in cameras which recognize the gestures via optical hand tracking. 

\begin{figure}[htbp]
\centerline{\includegraphics[width=0.5\textwidth]{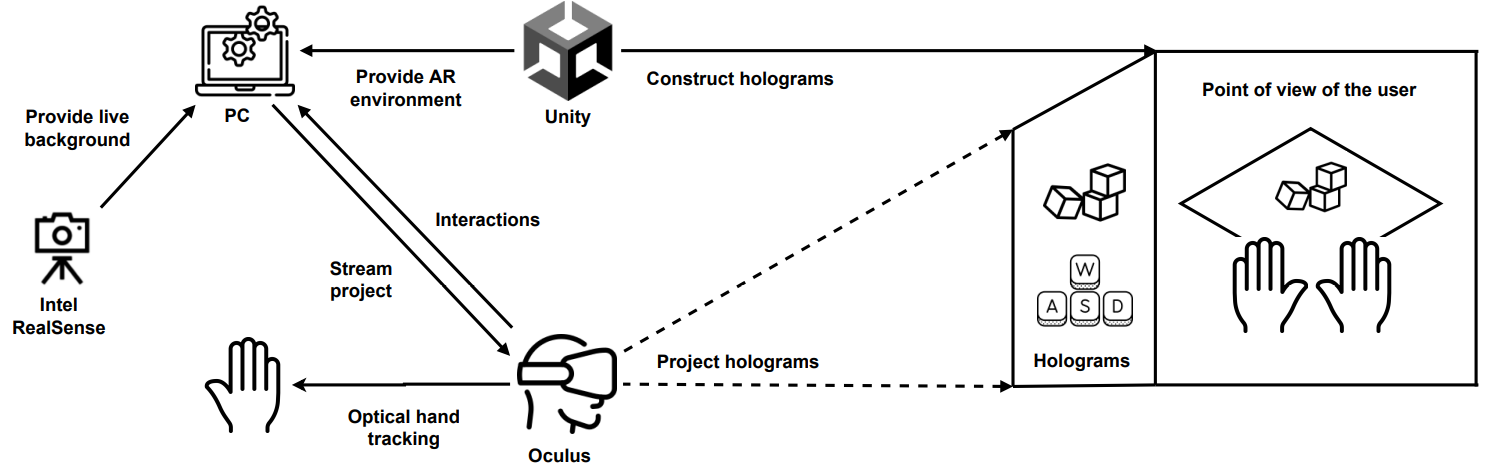}}
\caption{Schematic visualization of the system architecture using \acrshort{dim}.}
\label{fig:interactie2}
\end{figure}

Similar to Fig.~\ref{fig:interactie1}, a schematic visualization of the system architecture using \acrshort{dim} is presented in Fig.~\ref{fig:interactie2}. On the right, it is identical to the previous figure because for both interaction methods the same assignments are used. Different is that the user now wears an Oculus VR headset as \acrshort{hmd} due to an incompatibility between the HoloLens and the Manus gloves. The gloves are not used in this interaction method but to enable a comparison with the haptic gloves, \acrshort{dim} and \acrshort{hedim} use the same code. In contrast with the HoloLens, the project cannot be built to run standalone on the Oculus. Technically, this would be a possibility but then the Intel RealSense and the Manus gloves cannot be integrated. As both are essential for this research, it was required to stream the project with a USB cable. Consequently, the project executes on the personal computer. 

\begin{figure}[htbp]
\centerline{\includegraphics[width=0.5\textwidth]{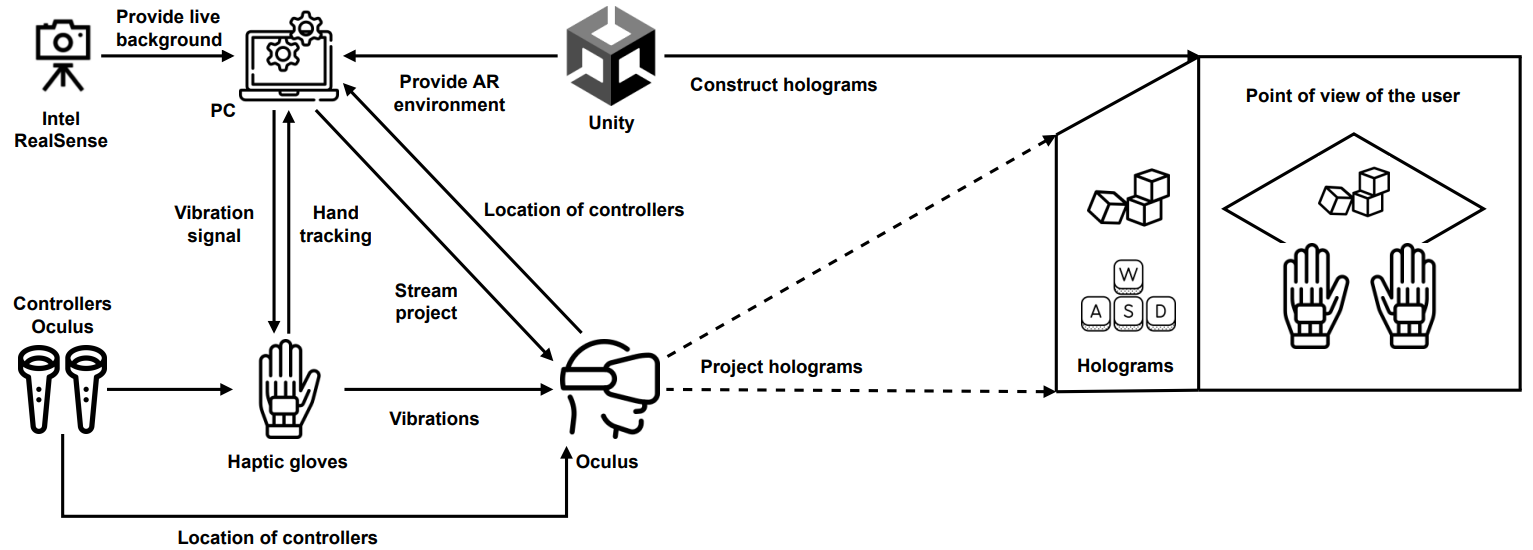}}
\caption{Schematic visualization of the system architecture using \acrshort{hedim}.}
\label{fig:interactie3}
\end{figure}

For the \acrshort{hedim}, a schematic visualization of the system architecture is displayed in Fig.~\ref{fig:interactie3}. This system architecture is similar to the one of \acrshort{dim} with the only difference being the haptic feedback. When the haptic gloves are used, they need the Oculus controllers to be attached to them. Because the gloves are not recognized as a person’s hand, optical hand tracking has become impossible. Therefore, the connected Oculus controllers will send the location of the hands via the \acrshort{hmd} to the personal computer. The gloves itself are wireless connected to the computer and they constantly send the shape of the users' hands to the computer. If a fingertip interacts with an \acrshort{ar} hologram, the computer will send a vibration signal that is felt by the user. 

\subsection{Implementation of the interaction methods}
\label{sec:Implementation}

As this research aims to improve standard interaction method, the test set-up of HoloLens could therefore not be optimized. Only guidelines given in Microsoft's official documentation were used. For further research and projects, it is essential to know in advance that Unity 2020 does not support targeting HoloLens (1st gen) anymore \cite{unityversion}. The \acrshort{hmd} remains supported in Unity 2019 LTS with Legacy Built-in XR for the full life cycle of Unity 2019 LTS through mid-2022.

The integrated keyboard was used for the standard interaction method of the HoloLens. An online notepad\cite{notepad} was used for showing and to make typing available for the user. Regarding the cube test set-up, the command design pattern was used to implement the interaction method of the HoloLens and the \acrshort{ar} environment was fully implemented in Unity. This interaction method is always performed from a distance, meaning without touching the virtual keyboard or cubes. The used Unity version was Unity2017.4.40f1 because it was the only version that allowed Holographic Remoting Player to be used. Without it being documented anywhere, in newer versions no connection can be made to the HoloLens without building the whole application. 

In contrast with the interaction method of the HoloLens, both the \acrshort{dim} and the \acrshort{hedim} implemented on the Oculus were optimized as much as possible. Firstly, both implementations are identical through reuse. The only difference is that for the \acrshort{hedim} the haptic gloves are activated. With this new interaction method, it is now possible to touch, grab and directly interact with virtual objects. When virtual object needs to be activated, moved or rotated the user needs to be close enough to the hologram. For the implementation of this interaction method, an observer design pattern was used where each fingertip was observed by the corresponding hand. Because the Oculus device had problems tracing the hands when the haptic gloves were worn, the prefab was slightly modified for the haptic version. The standard optical hand tracking was changed by the built-in hand tracking of the gloves. It should be noted that apart from this nothing has changed in the way the interaction method works.

\section{Methodology}
\label{sec:Assignments}

To obtain two objective performance parameters, the time and score of the user per assignment for each interaction method is measured. This section will discuss in detail how the subjects are evaluated on these parameters. For the subjective parameters a questionnaire using Likert scale has been used.

\begin{figure}[htbp]
\centerline{\includegraphics[width=0.5\textwidth]{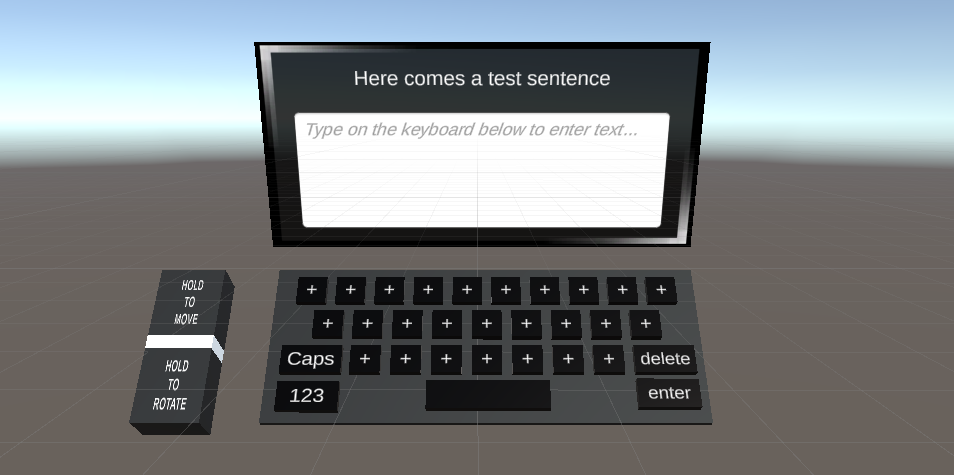}}
\caption{Visualization of the set-up of the keyboard test. (Keys are loaded by runtime)}
\label{fig:keyboardset}
\end{figure}

For the keyboard test, each subject was requested to type a random sentence that complied to the following self-imposed rules. First, the sentence had to consist of a total of 30 characters. Second, it had to contain three special characters that were not allowed to be in a row. Third, it also had to include three numbers that had to be in a row. The remaining 24 characters had to contain five capital letters that were also not allowed to occur consecutively. The space character was not considered as a special character and had therefore no additional conditions. Finally, the language of the sentences was always English. To give each user the same difficulty, 3 fixed sentences were chosen. These were randomly used for a different interaction method for each user. We assembled the following sentences:
\begin{center}
    \#HelloWorld\textless Test No Error-404 \\
	@Ghent is for me ToP-100 CitY\textasciicircum \\
	\textgreater123 i LoVe auGmented-RealitY!  \\
\end{center}
After exactly 4 minutes the test would be stopped. for each error in the sentence or for each missing character, an error score was counted. This error score was then added to the number of times delete was pressed during the assignment to get the total number of errors. The error score for this assignment is therefore the number of mistakes made. This means that a lower error score is better than a higher one. A visualization of the set-up of the keyboard test is presented in Fig.~\ref{fig:keyboardset}.

\begin{figure}[htbp]
\centerline{\includegraphics[width=0.5\textwidth]{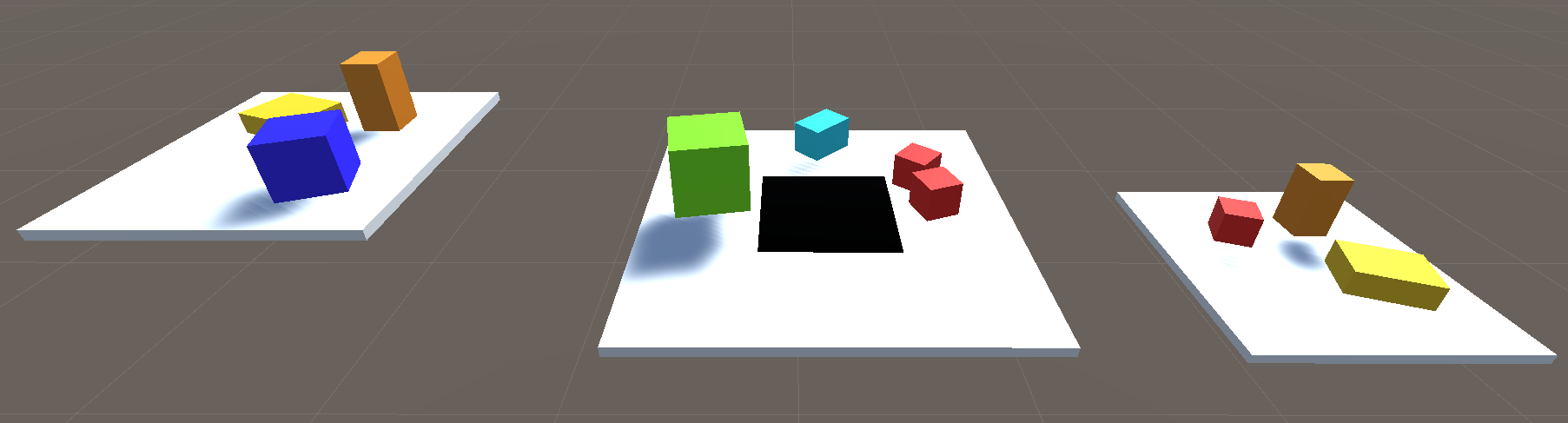}}
\caption{Visualization of the set-up of the cube test.}
\label{fig:cubeset}
\end{figure}

After successfully completing the assignment on the keyboard, the subjects were presented with the cube test. The goal of this test was to make a tower of six blocks high in the center platform. Blocks could fall when moving from one of the two other platforms. When this happened, that cube was permanently gone. This was deliberately done to prevent the user from building indefinitely until time ran out. Additionally, it was also intended to give users a real sense of grip while holding a cube and create frustration when dropping one. A visualization of the set-up of the cube test is presented in Fig.~\ref{fig:cubeset}. The score was determined by subtracting the height (in number of blocks) of the user's largest tower from the requested height (i.e. six blocks). Furthermore, two times the number of fallen cubes was added to this. This gives the following formula:

\begin{center}

Score = 6 - (tallest tower of the user) + 2 * (fallen cubes)

\end{center}
The score for this assignment is therefore again the number of mistakes made. It was decided to give more weight to the fallen blocks in order to better deduce how well the user could operate a block. This means both the grip while holding a block, but also the precise placement of the block on top of the tower in the three-dimensional space because blocks can still fall if the tower is poorly built.

\section{Discussion of results}
\label{sec:discussion}
\begin{figure}[htbp]
\centerline{\includegraphics[width=0.5\textwidth]{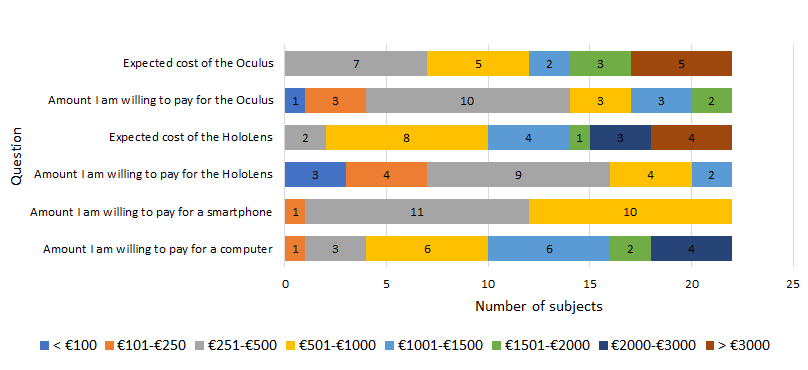}}
\caption{Bar chart containing the financial preferences of the subjects.}
\label{fig:money}
\end{figure}

\begin{figure}[htbp]
\centerline{\includegraphics[width=0.5\textwidth]{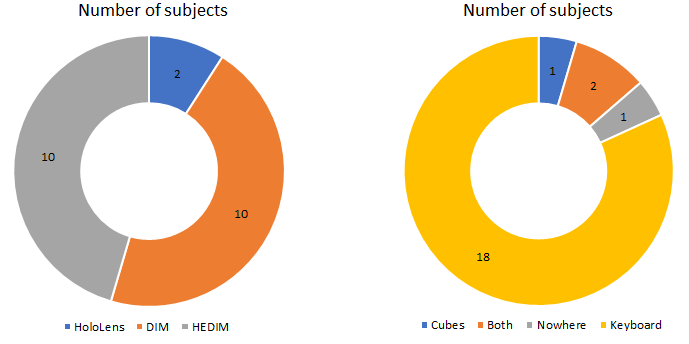}}
\caption{Doughnut chart of the most preferred interaction method (left) and doughnut chart of the assignment where the haptic gloves were most useful (right)}
\label{fig:endresult}
\end{figure}

This section discusses results of this research. First, in Table~\ref{tab:score} a summary including statistical values of the error score of the tests is presented. Thereafter, in Table \ref{tab:time} the similar statistical values are shown for the time parameter. Next, a bar chart containing the financial preferences of the subjects is displayed in Fig.~\ref{fig:money}. Finally, Fig.~\ref{fig:endresult} contains a doughnut chart of the most preferred interaction method and a doughnut chart of the assignment where the haptic gloves were most useful.

Prior to discussing the results, it is valuable to analyze the test group. For this study, a minimum of 20 people was envisioned and in total 22 subjects were actually tested. It should be noted that all test subjects participated voluntarily without compensation. Furthermore, no public call was made to find participants. As a result, everyone who participated was either an acquaintance or an employee of the research group. Nevertheless, none of the subjects knew in advance the purpose of the study to ensure objectivity. 

The age of the participants ranged from 16 years to 50 years old. An average age of 25.95 years was obtained with a standard deviation of 8.82 years. Similarly, the gender of the participants was well distributed. The research included 10 women and 12 men, which corresponds to a distribution of 45.45\% women and 54.55\% men. 

\begin{table}[htbp]
	\caption{Summary statistical values of the error score with the best (lowest) mean score per assignment in bold.}
	\label{tab:score}
	\begin{tabular}{ c c | r r r}
		\toprule
		Interaction method & Task & Mean & Std dev & Variance\\
		\midrule
		HoloLens & K & 8.23 & 7.12 & 50.63 \\
		HoloLens & C & 3.32 & 3.53 & 12.49 \\
		\hline
		\acrshort{dim} & K & 29.91 & 22.94 & 526.45 \\
		\acrshort{dim} & C & 2.59 & 3.81 & 14.51 \\
		\hline
		\acrshort{hedim} & K & \textbf{5.77} & 7.35 & 53.99 \\
		\acrshort{hedim} & C & \textbf{1.86} & 2.34 & 5.48 \\

		\bottomrule
	\end{tabular}
\end{table}

The first thing that stands out in Table~\ref{tab:score} is that the mean score of the \acrshort{hedim} is the lowest, thus the best, for both the assignments. For the cube assignment, the same method without the haptic gloves, i.e. \acrshort{dim}, still performed better than the standard interaction method of the HoloLens.  For the keyboard assignment, a significant higher mean value can be observed. When normal distribution is assumed, for the keyboard assignment, a user using the \acrshort{hedim} has 57.77\% chance to score better than the standard interaction method of the HoloLens and 84.28\% chance to score better than the \acrshort{dim}. Similarly for the cube test, users have 63.5\% chance to score better than using the standard interaction method and 56.49\% chance to score better than the \acrshort{dim}.

\begin{table}[htbp]
	\caption{Summary statistical values of the time (in seconds) with the fastest mean time per assignment in bold.}
	\label{tab:time}
	\begin{tabular}{ c c | r r r}
		\toprule
		Interaction method & Task & Mean & Std dev & Variance\\
		\midrule
		HoloLens & K  & 182.23 & 53.90 & 2905.36 \\
		HoloLens & C & 169.64 & 64.36 & 4142.60 \\
		\hline
		\acrshort{dim} & K & 158.14 & 50.57 & 2557.66 \\
		\acrshort{dim} & C & \textbf{62.00} & 37.29 & 1390.82 \\
		\hline
		\acrshort{hedim} & K & \textbf{81.27} & 29.25 & 855.74 \\
		\acrshort{hedim} & C & 103.64 & 56.46 & 3187.32 \\

		\bottomrule
	\end{tabular}
\end{table}

By analyzing Table \ref{tab:time}, it cannot be easily concluded which interaction method performs best. This is because \acrshort{dim} and \acrshort{hedim} each perform superior on a different assignment. On the other hand, the standard interaction method of the HoloLens has the highest time for both tests. Consequently, this is the worst interaction method in terms of speed. When normal distribution is assumed, for the keyboard assignment, a user using the \acrshort{hedim} has 90.75\% chance to finish earlier than the \acrshort{dim} and 95.10\% chance to finish earlier than standard interaction method. For the cube test, a user has a higher chance to finish early with \acrshort{dim}. They have 73.08\% chance to score a better time than with \acrshort{hedim} and 92.60\% chance to score better than with the standard interaction method.

By analyzing the graph in Fig.~/ref{fig:money}, it can be observed that 72.72\% is not willing to spend more than 1,500 euros on a computer and 100\% is not willing to spend more than 1,000 euros on a mobile phone. If the hypothesis is correct, the users are not willing to pay more than 1.000 euros on an \acrshort{hmd} and definitely not more than €1.500. As the HoloLens is the only true \acrshort{ar} device, it is first analyzed. From all test subjects, 100.00\% are not willing to pay more than 1.500 euros and 90.91\% are not willing to pay a maximum of 1.000 euros for it. Secondly, 90.91\% do not want to pay more than 1.500 euros for a VR HMD and 77.27\% have a limit of 1,000 euros. From this, it can be concluded that the majority do not want to spend more than they currently spend on a cell phone or computer.

The HoloLens (First generation) was sold in 2016 for 3,000 dollars for a developer's kit and 5,000 for a commercial suite \cite{prices1}. Nowadays, this is respectively 2922.97 euros and 4871.63 euros. The price of the HoloLens 2 is 3500 dollars, meaning 3410.14 euros at the moment. The Oculus Meta Quest is available between 449.99 and 549.99 euros depending on desired storage capacity \cite{prices2}. Since solely price ranges were asked and the price of the HoloLens depends on the version, both the price range 2000-3000 euros and above 3000 euros are approved. This results in 7 out 22 subjects or 31.82\% having correctly estimated the price. Similarly, for the Oculus Quest both the price range 251-500 euros and 501-1000 euros are approved. As a result, 13 of the 22 people or 59.09\% were able to indicate the correct price range. Due to the large distance between the boundaries of the price ranges and because only 45.45\% of the total answers were correct, it is concluded that users are not able to estimate the costs correctly for an \acrshort{hmd}.

Before a final conclusion can be made whether the haptic feedback and the new interaction method improves the standard interaction method of the HoloLens, all participants were asked which interaction method they preferred for daily use, and in which test they found the haptic gloves the most useful. The result of their most desired interaction method is visible in Fig.~\ref{fig:endresult}. The name of the \acrshort{hmd}s was not used so that the users could only take into account the interaction method and were not biased by the \acrshort{hmd}s. Additionally, in Fig.~\ref{fig:endresult}, it is displayed for which assignment the users found the haptic gloves most useful.

It can be observed that the test subjects have no general preference in favor or against haptic feedback. Nevertheless, the standard HoloLens interaction method is clearly not preferred. This allows the study to conclude that both the \acrshort{dim} and the \acrshort{hedim} exceeds the standard method among users. When looking at the preferences when only considering haptic feedback, the following result is obtained: one user prefers the cube assignment, 16 users prefer the keyboard, three users find both applications useful with haptic feedback and two users do not find the haptic gloves useful anywhere. Users find the keyboard the most useful application for haptic gloves. Additionally, \acrshort{hedim} scored the lowest mean time necessary for the keyboard task. In contrast, for the cube assignment it was the \acrshort{dim} with the best time. However, one may wonder whether gloves of 3,999 euros are worth buying if only used as a keyboard. This research concludes that it is clearly not worth the financial cost. This conclusion has also been confirmed by the Financial Hypothesis in which it was concluded that people do not want to spend more than they currently spend on a cell phone or computer solely on the \acrshort{hmd} and, therefore, certainly do not want to pay an extra 3,999 euros.

\section{Conclusion}
First, after analyzing the results, it was first concluded that the user experience with all \acrshort{hmd}s were equal and that the switch in \acrshort{hmd}s was without influence on the result of the further discussed conclusions. Second, this research deduced from the gathered data that users have a slight preference for using the \acrshort{dim} in general situations. For specific situations, users preferred using the \acrshort{dim} for the cube assignment and using the \acrshort{hedim} for the keyboard task. This subjective perspective of the users could be supported by the objective time measurements which scored the best for the cube assignment when using the \acrshort{dim} and for the keyboard when using the \acrshort{hedim}. For the score parameter, it was concluded that using the \acrshort{hedim} obtained the best result for both assignments. It should be noted that despite the differences in the \acrshort{dim} and the \acrshort{hedim}, both outscored the standard interaction method of the HoloLens each time. Last, this study determined that the majority of users do not want to spend more than they currently spend on a cell phone or computer for an \acrshort{hmd} and users are not able to estimate the costs correctly of an \acrshort{hmd}.

This research concludes as an end result that the \acrshort{dim} and the \acrshort{hedim} have proven to improve the standard method of the HoloLens. Additionally, it has been established that haptic feedback enhances the \acrshort{dim} in certain situations. Nevertheless, in this study, the haptic gloves only make a small improvement that it is not worth the financial cost. As future work, we recognize that \acrshort{dim} could still be improved in comparison with the haptic feedback by adding software enhancements. As a result, the research concludes that, at this moment, haptic feedback is useful for high-end personalized applications, but not yet for daily use. 

\bibliography{conference_101719}

\end{document}

%% file: acronyms.tex

\newacronym{ar}{AR}{Augmented Reality}
\newacronym{hmd}{HMD}{Head-Mounted Display}
\newacronym{vr}{VR}{Virtual Reality}
\newacronym{mr}{MR}{Mixed Reality}
\newacronym{xr}{XR}{eXtended Reality}
\newacronym{av}{AV}{Augmented Virtuality}
\newacronym{sr}{SR}{Simulated Reality}
\newacronym{vhmr}{VHMR}{Visio-Haptic Mixed Reality}
\newacronym{nasa}{NASA}{National Aeronautics and Space Administration}
\newacronym{marta}{MARTA}{Mobile Augmented Reality Technical Assistance}
\newacronym{dof}{DoF}{Degrees of Freedom}
\newacronym{fps}{FPS}{Frames Per Second}
\newacronym{rfid}{RFID}{Radio-Frequency IDentification}
\newacronym{qos}{QoS}{Quality-of-Service}
\newacronym{qoe}{QoE}{Quality-of-Experience}
\newacronym{jet}{JET}{Joint European Torus}
\newacronym{afrl}{AFRL}{United States Air Force Research Laboratory}
\newacronym{imu}{IMU}{Inertial Measurement Unit}
\newacronym{dll}{dll}{Dynamic-Link Library}
\newacronym{fov}{FoV}{Field of View}
\newacronym{hpu}{HPU}{Holographic Processing Unit}
\newacronym{srg}{SRG}{Surface Relief Grating}
\newacronym{vhg}{VHG}{Volumetric Holographic Grating}
\newacronym{lra}{LRA}{Linear Resonant Actuators}
\newacronym{ipd}{IPD}{Inter-Pupillary Distance}
\newacronym{gdpr}{GDPR}{General Data Protection Regulation}
\newacronym{ssq}{SSQ}{Simulator Sickness Questionnaire}
\newacronym{sdgs}{SDGs}{Sustainable Development Goals}
\newacronym{dim}{DIM}{Direct Interaction Method}
\newacronym{hedim}{HEDIM}{Haptic Enhanced Direct Interaction Method}
\newacronym{idlab}{IDLAB}{Internet Technology And Data Science Lab}
\newacronym{mreq}{MREQ}{Mixed Reality Experience Questionnaire}